\documentclass{article}

\usepackage{arxiv}

\usepackage[utf8]{inputenc} % allow utf-8 input
\usepackage[T1]{fontenc}    % use 8-bit T1 fonts
\usepackage[colorlinks=True]{hyperref}       % hyperlinks
\usepackage{url}            % simple URL typesetting
\usepackage{booktabs}       % professional-quality tables
\usepackage{amsfonts}       % blackboard math symbols
\usepackage{nicefrac}       % compact symbols for 1/2, etc.
\usepackage{microtype}      % microtypography
\usepackage{lipsum}
\usepackage{listings}

\usepackage{xcolor}
\usepackage{amsmath}
\usepackage{bm}
\usepackage{subcaption}
\usepackage{multirow}
\usepackage{makecell}
\usepackage{graphicx}
\usepackage{appendix}
\usepackage{grffile}

\DeclareMathOperator*{\argmin}{argmin}  % in your preamble 

\newenvironment{myfont}{\fontfamily{pcr}\selectfont}{\par}
\DeclareTextFontCommand{\textmyfont}{\myfont}

\title{Adversarial Attack and Defense Strategies for Deep Speaker Recognition Systems}

\author{
  Arindam Jati$^\dagger$\thanks{Authors contributed equally.} \\
  \texttt{jati@usc.edu} \\
   \And
  Chin-Cheng Hsu$^{\dagger*}$ \\
  \texttt{chincheh@usc.edu} \\
   \And
  Monisankha Pal$^\dagger$ \\
  \texttt{mp\_323@usc.edu} \\
   \AND
  Raghuveer Peri$^\dagger$ \\
  \texttt{rperi@usc.edu}
  \And
  Wael AbdAlmageed$^{\dagger\S}$ \\
  \texttt{wamageed@isi.edu}
  \And
   Shrikanth Narayanan$^{\dagger\S}$ \\
   \texttt{shri@sipi.usc.edu} \\
  \AND
  \\
  $^\dagger$Electrical and Computer Engineering, University of Southern California (USC), Los Angeles, CA, USA\\
  $^\S$USC Information Sciences Institute, Marina del Rey, CA, USA
}

\begin{document}
\maketitle

\newcommand\blfootnote[1]{%
  \begingroup
  \renewcommand\thefootnote{}\footnote{#1}%
  \addtocounter{footnote}{-1}%
  \endgroup
}

\begin{abstract}
Robust speaker recognition, including in the presence of malicious attacks, is becoming increasingly important and essential, especially due to the proliferation of several smart speakers and personal agents that interact with an individual's voice commands to perform diverse, and even sensitive tasks. Adversarial attack is a recently revived domain which is shown to be effective in breaking deep neural network-based classifiers, specifically, by forcing them to change their posterior distribution by only perturbing the input samples by a very small amount. Although, significant progress in this realm has been made in the computer vision domain, advances within speaker recognition is still limited. The present expository paper considers several state-of-the-art adversarial attacks to a deep speaker recognition system, employing strong defense methods as countermeasures, and reporting on several ablation studies to obtain a comprehensive understanding of the problem. The experiments show that the speaker recognition systems are vulnerable to adversarial attacks, and the strongest attacks can reduce the accuracy of the system from 94\% to even 0\%. 
The study also compares the performances of the employed defense methods in detail, and finds adversarial training based on Projected Gradient Descent (PGD) to be the best defense method in our setting. We hope that the experiments presented in this paper provide baselines that can be useful for the research community interested in further studying adversarial robustness of speaker recognition systems.
\end{abstract}

%shortcuts
\newcommand{\eg}{\textit{e.g.,\ }}
\newcommand{\etc}{\textit{etc.\@}}
\newcommand{\vs}{\textit{vs.\@ }}
\newcommand{\ie}{\textit{i.e.,\ }}
\newcommand{\etal}{\textit{et al. \@}}
\newcommand{\cf}{\textit{cf.,\ }}

\newcommand{\adv}{\boldsymbol{\eta}}
\newcommand{\x}{\boldsymbol{x}}
\newcommand{\param}{\boldsymbol{\theta}}
\newcommand{\noisyx}{\widetilde{\boldsymbol{x}}}

\newcommand{\X}{\mathcal{X}}
\newcommand{\Y}{\mathcal{Y}}
\newcommand{\D}{\mathcal{D}}

\newcommand{\norm}[1]{\left\lVert#1\right\rVert}

% keywords can be removed

\section{Introduction}
%\paragraph{Adversarial attacks.}
Deep learning models are recently found to be vulnerable to \textit{adversarial attacks}~\cite{szegedy2013intriguing,biggio2013evasion}  where the attacker potentially discovers blind spots in the model, and crafts \textit{adversarial samples} that are only slightly different from the original samples, rendering the trained model fail to correctly classify them or even to perform any other inference task on them.
Over the last few years, several researchers have devoted significant effort in devising novel adversarial attack algorithms~\cite{goodfellow2014explaining,madry2018towards,papernot2016limitations,carlini2017towards}, proposing defensive countermeasures to gain robustness~\cite{goodfellow2014explaining,madry2018towards}, and demonstrating exploratory analyses~\cite{carlini2017towards,athalye2018obfuscated,carlini2019evaluating}.

\paragraph{Adversarial attack on speech processing systems.} 
With the rapid increase in the incorporation of Deep Neural Networks~(DNN) within speech processing applications like Automatic Speech Recognition~(ASR)~\cite{chan2016listen,Audhkhasi2017}, speaker recognition~\cite{hansen2015speaker,Chung2018,snyder2018x,jati2019neural}, and speech emotion and behavior studies~\cite{narayanan2013behavioral,huang2017deep}, it is becoming essential to study the probable weaknesses of the employed models in the presence of adversarial attacks.
In~\cite{carlini2018audio}, the authors have shown that it is possible to achieve even $100\%$ \textit{success rate} in attacking deep ASR systems.
In~\cite{qin2019imperceptible} the authors have successfully generated imperceptible (to humans) adversarial audio samples while retaining high attack success rate.
These studies highlight the vulnerability of deep ASR models against adversarial attacks.

\paragraph{Adversarial attack on speaker recognition systems.}
Speaker recognition models are being widely employed in several applications including smart speakers and personal digital assistants~\cite{wan2018generalized,hansen2015speaker}, bio-metric systems~\cite{nucci2012hierarchical}, and forensics~\cite{becker2008forensic}.
Therefore, having robust speaker recognition models that are not susceptible to adversarial perturbation is an important requirement.
However, speaker recognition models have \textit{not} been investigated extensively in the presence of adversarial attacks.
Some initial work can be found in the literature (please refer to Section~\ref{sec:Related work}), but a detailed analysis of \textit{white box} attacks (will be discussed in Section~\ref{subsec:Adversarial attack}) with state-of-the art attack algorithms is difficult to find.
Moreover, to the best of our knowledge, effective defensive countermeasures for those attacks have \textit{not} been proposed.
The present work aims to address these issues in particular.

\paragraph{Contributions.}
This paper focuses on adversarial attacks and possible countermeasures for deep speaker recognition systems, with
 the following contributions.
\begin{itemize}
    \item In contrast to previous works in this field (discussed in Section~\ref{sec:Related work}), we perform adversarial attack directly on the time domain speech signal (and \textit{not} on the spectrogram), which is more realistic in real-life scenarios.
    \item We provide an extensive analysis of the effect of multiple state-of-the-art white box adversarial attacks on a DNN-based speaker recognition model.
    \item We propose multiple defensive countermeasures for the deep speaker recognition system, and analyze their performance.
    \item We perform \textit{transferability analysis}~\cite{carlini2019evaluating} to investigate how adversarial speech crafted with a particular model can also be harmful to a different model.
    \item We present various ablation studies (\eg varying the strength of the attack, measuring signal-to-noise ratio~(SNR) and perceptibility of the adversarial speech samples \etc) that might be helpful to gain a comprehensive understanding of the problem.
    \item We share ready-to-run software implementation\footnote{Source codes are available at \url{https://github.com/usc-sail/gard-adversarial-speaker-id}} of the present work toward supporting reproducibility and further research.
\end{itemize}
We aim to set baselines in the present exposition study, and hope it can help the community interested to continue further research in this domain.

\paragraph{Paper outline.}
The rest of the paper is organized as follows. In Section~\ref{sec:Preliminary}, we provide preliminaries about speaker recognition and adversarial attack. In Section~\ref{sec:Related work}, we highlight the related work. The adversarial attack algorithms and defense strategies are introduced in Section~\ref{sec:Methodology}. Experimental setting and results are described in Section~\ref{sec:Experimental setting} and Section~\ref{sec:Results and discussions}, respectively. Finally, conclusions and future directions are provided in Section~\ref{sec:Conclusion and future directions}.

\section{Preliminary} \label{sec:Preliminary}
\subsection{Speaker recognition systems}
Speaker recognition systems can be developed either for identification or verification~\cite{hansen2015speaker} of individuals from their speech.
In a \textit{closed set} speaker identification scenario~\cite{hansen2015speaker,jati2019neural}, we are provided with train and test utterances from a set of unique speakers.
The task is to train a model that, given a test utterance, can classify it to one of the training speakers.
Speaker verification~\cite{snyder2018x,Chung2018}, on the other hand, is an \textit{open set} problem.
The task is to verify whether a test utterance claiming a particular speaker's identity is actually spoken by that speaker (whose enrolment utterance is available beforehand).
The training data in the latter case, is generally utterances from a mutually exclusive set of speakers.

Although, speaker verification differs from speaker identification during the testing phase, most of the recent state-of-the-art speaker verification systems~\cite{snyder2018x,Nagrani2017,Chung2018,snyder2017deep} are trained with the objective of learning to classify the set of training speakers.
In other words, these models are trained with a cross-entropy objective over the unique set of training speakers (\ie similar to a speaker identification scenario). 

Formally, if $\boldsymbol{x} \in \mathbb{R}^D$ denotes a time domain audio sample with speaker label $y$, then learning a speaker identifier model is generally done through Empirical Risk Minimization (ERM)~\cite{madry2018towards}:
\begin{equation}
\label{eq:cross-entropy}
    \argmin_{\param} \quad \mathbb{E}_{ \left( \x ,y \right) \sim \D} \left[ L\left( \x, y, \param \right) \right]
\end{equation}
where, $L(\cdot)$ is the cross-entropy objective, and $\param$ denotes the set of trainable parameters of the DNN. 

An intermediate representation of the trained DNN model might be subsequently extracted as a \textit{speaker embedding}~\cite{snyder2018x} which is expected to carry speaker-specific information.
The speaker embeddings are then utilized for verification purposes.
Because of this widespread use, in this study, we work with a closed set speaker identification (or classification) model.
The findings of this study can motivate future research on open set speaker verification task (see Section~\ref{sec:Conclusion and future directions} for future directions).

\subsection{Adversarial attack}\label{subsec:Adversarial attack}
Given an audio sample $\boldsymbol{x}$, an adversarial attack generates a perturbed signal given by
\begin{equation}
    \noisyx = \x + \adv \quad \text{such that} \quad \norm{\adv}_p < \epsilon
\end{equation}
with the goal of forcing the classifier to produce erroneous output for $\noisyx$.
In other words, if $\x$ has a true label $y$, then the attacker forces the classifier to produce $\widetilde{y} \ne y$ for the perturbed sample $\noisyx$.
In this paper, we will focus on $l_\infty$ and $l_2$ norms which are most widely employed in the literature.

\subsubsection{Threat model}
We explore \textit{white-box}~\cite{carlini2019evaluating} attack in this study.
This assumes that the attacker has complete knowledge of the model architecture, parameters, loss functions, and gradients.
We adopt this stronger form of attack (compared to black-box attack~\cite{carlini2019evaluating}) because it does not assume that any part of the model can be kept hidden from the attacker, and it is the most frequently employed threat model in the adversarial attack literature~\cite{goodfellow2014explaining,madry2018towards,papernot2016limitations,carlini2017towards}.

Adversarial attack can be \textit{targeted} or \textit{untargeted}~\cite{carlini2019evaluating}.
An untargeted attack only forces the model to generate erroneous outputs, whereas, a targeted attack forces the model to predict a target class which is different from the true class.
We perform \textit{untargeted attacks} in this study, and leave the targeted attack for future study (see Section~\ref{sec:Conclusion and future directions}).

\subsubsection{Transferability}
Although most of the experiments in this paper are with white box attacks, we study the transferability of adversarial samples in Section~\ref{subsec:Ablation study 3: Transferability analysis}, which gives us a notion of performance during a black box attack as well. The transferability test~\cite{carlini2019evaluating,papernot2016transferability} evaluates the vulnerability of a \textit{target} model against the adversarial samples generated with a \textit{source} model. The attacker has full knowledge about the source model, but no or limited knowledge about the target model (for example, knowledge about the fact that both source and target have convolutional layers). The goal of the attacker is to generate adversarial samples (with the source model) in such a way that they ``transfer well'' to the target model, \ie those samples also make the target model vulnerable.

\section{Related Work}\label{sec:Related work}
This section describes key previous work on adversarial attack and defense methods proposed for speaker recognition systems. 
\begin{itemize}
    \item Li \etal~\cite{li2020adversarial} showed that an i-vector~\cite{dehak2010front} based speaker verification system is susceptible to adversarial attacks, and the adversarial samples generated with the i-vector system also transfer well to a DNN-based x-vector~\cite{snyder2018x} system\footnote{i-vectors have been the state-of-the-art in speaker verification for a decade until DNN-based x-vectors were shown to outperform them~\cite{snyder2017deep,snyder2018x}.}. The attack was performed on the feature space (and not directly on the time domain speech signal), and with only the Fast Gradient Sign Method~(FGSM)~\cite{goodfellow2014explaining} (will be further discussed in Section~\ref{sec:Methodology}) was investigated for that purpose. Moreover, no defense method was proposed.
    \item Kreuk \etal~\cite{kreuk2018fooling} demonstrated the vulnerability of an end-to-end DNN-based speaker verification system to FGSM attack. The attack was done on the feature space, and the authors discovered cross-feature transferability of the adversarial samples. No defense method was proposed in the paper.
    \item Chen \etal~\cite{chen2019real} proposed the Natural Evolution Strategy~(NES) based adversarial sample generation procedure, and successfully attacked a GMM-UBM system\footnote{GMM-UBM stands for Gaussian Mixture Model-Universal Background Model, a classical model in speaker recognition~\cite{reynolds2000speaker}.} and i-vector based speaker recognition systems. They found impressive attack success rate with their proposed method. However, the authors did not attack more recent DNN-based speaker recognition frameworks which are shown to have state-of-the-art performances. Moreover, the test set involved in their experiments only included $5$ speakers (TABLE I of~\cite{chen2019real}), and thus, an extensive study with a much higher number of test speakers is still needed.
    \item Wang \etal~\cite{wang2019adversarial} proposed adversarial regularization based defense methods using FGSM and Local Distributional Smoothness~(LDS)~\cite{miyato2015distributional} techniques. The proposed method was shown to improve the performance of a speaker verification system, but only FGSM was employed as the attack algorithm, and similar to most of the above methods, the attack was performed on the feature space and not on the time domain audio.
\end{itemize}
In summary, although these studies represent important initial efforts on adversarial attacks on speaker recognition system, many technical questions still remain to be addressed. Limitations include consideration of primarily feature space attacks~\cite{li2020adversarial,kreuk2018fooling,wang2019adversarial} (and not time domain), limited number of attack algorithms~\cite{li2020adversarial,kreuk2018fooling,chen2019real,wang2019adversarial}, limited number of speakers in the test set~\cite{chen2019real}, and no or limited number of defense methods~\cite{li2020adversarial,kreuk2018fooling,chen2019real}. 
The present exposition study aims to address some of these limitations by reporting extensive experimental analysis,  ablation studies, and by proposing and evaluating various defense methods.

\section{Attack and Defense Algorithms}
\label{sec:Methodology}
\subsection{Attack algorithms}

A group of gradient-based attack algorithms tries to maximize the loss function by finding a suitable perturbation which lies inside the $l_p$-ball around $\x$. Formally,
\begin{equation}
    \max_{\adv:\norm{\adv}_p < \epsilon} \quad  L\left( \x+\adv, y, \param \right).
\end{equation}
A different group of algorithms aims at decreasing the posterior of the true output class, and increasing the posterior of the most confusing wrong class.
Here we present the attack algorithms we employ in our study. 

\paragraph{Fast Gradient Sign Method (FGSM).} 
Goodfellow \etal~\cite{goodfellow2014explaining} proposed this computationally efficient one-step $l_\infty$ attack to generate adversarial samples by only using the sign of the gradient function, and moving in the direction of gradient to increase the loss:
\begin{equation}
    \label{eq:fgsm}
    \noisyx = \x + \epsilon \text{ sign} \left( \nabla_{\x}L\left( \x, y, \param \right) \right).
\end{equation}

\paragraph{Projected Gradient Descent (PGD).} 
Madry \etal~\cite{madry2018towards} proposed a more generalized version with iterative gradient based $l_\infty$ attack:
\begin{equation}
    \label{eq:pgd}
    %\noisyx_{i+1} = \underset{\norm{\x-\noisyx_i}_\infty < \epsilon}{\Pi}
    %\left[ \noisyx_i + \alpha \text{ sign} \left( \nabla_{\x}L\left( \x, y, \param \right) \right) \right],
    \noisyx_{i+1} = \Pi_{\x + \mathcal{S}}
    \left[ \noisyx_i + \alpha \text{ sign} \left( \nabla_{\x}L\left( \x, y, \param \right) \right) \right],
\end{equation}
where, $\alpha$ is the step size of the gradient descent update,
$\x+\mathcal{S}$ is the set of allowed perturbations \ie the $l_\infty$-ball around $\x$, and $\Pi_{\x + \mathcal{S}}$ denotes the constrained projection operation in a standard PGD optimization algorithm.
PGD is run for a fixed number of maximum iterations, $T$. Throughout the text, we will denote PGD run for $T$ iterations by ``PGD-$T$''. 

\paragraph{Carlini and Wagner attack (Carlini $l_2$ and Carlini $l_\infty$).} 
Carlini and Wagner~\cite{carlini2017towards} defined the general methodology of their attack by
\begin{equation}
    \begin{aligned}
        \text{minimize} &\quad \norm{\adv}_p + c \cdot g(\noisyx) \\
        \text{such that} &\quad \noisyx \in [0,1].
    \end{aligned}
\end{equation}
Here, $g(\cdot)$ defines the objective function given by
\begin{equation}
    g(\noisyx) = \left[ Z(\noisyx)_t - \max_{j \ne t} \left( Z(\noisyx)_j\right) + \delta \right]_+
\end{equation}
where, $Z(\cdot)$ is the output vector containing posterior probabilities for all the classes, $t$ denotes the output node corresponding to the true class $y$, $\delta$ is the confidence margin parameter, and $[\cdot]_+$ denotes the $\textrm{max}(\cdot, 0)$ function.
Intuitively, the attack tries to maximize the posterior probability of a class that is \textit{not} the true class of $\x$, but has the highest posterior among all the wrong classes. 
The norm can be either $l_2$ or $l_\infty$. For Carlini $l_\infty$ attack, the minimization of $\norm{\adv}_\infty$ is not straightforward due to non-differentiability, and an iterative procedure is employed in~\cite{carlini2017towards}\footnote{We suggest the readers to refer to~\cite{carlini2017towards} for detailed information about the iterative workaround for $l_\infty$ attack, and also for choosing the values for the weight parameter, $c$.}.

\subsection{Defense algorithms} \label{subsec:Defense algorithms}
% We investigated three defense mechanisms in our experiments. 

\paragraph{Adversarial training.} 
%OLD one written by Jeremy
% First, we tested FGSM-based adversarial training,
% in which we trained the model with clean speech samples $\x$ and adversarial samples $\noisyx$ that were generated using FGSM:
% \begin{equation}
% \label{eq:fgsm}
%     \noisyx_{i + 1} := \noisyx_i + \adv = \noisyx_i + \epsilon ~\textrm{sign}( \nabla_{\noisyx_i} L(\noisyx_i, y;\param) ),
% \end{equation}
% where 
% $\noisyx_0 := \x$,
% $i \in \{0\} \cup \mathcal{N}$ denotes the $i$-th iteration,
% $\adv$ denotes the injected adversarial noise, and 
% $\epsilon$ denotes a small constant (treated as a hyper-parameter in our experiments).
% In FGSM-based adversarial training, $i=1$ and 
% the loss function $L$ in \eqref{eq:cross-entropy} is augmented with an additional weighted loss from the adversarial sample, resulting in overall loss function $L_G$:
% \begin{equation}
%     L_G(\x, \noisyx, y;\param) ) := L(\x, y;\param) ) + \alpha ~L(\noisyx, y;\param) ),
% \end{equation}
% where $\alpha$ is the weight of the adversarial training.

The intuition here is to train the model on adversarial samples generated by a certain adversarial attack.
The adversarial samples are generated online using the training data and the current model parameters.
Madry \etal~\cite{madry2018towards} introduced the generalized notion of adversarial training by a \textit{mini-max optimization} given by:
\begin{equation}
    \argmin_{\param} \quad \mathbb{E}_{(\x,y) \sim \D} \left[ \max_{\adv:\norm{\adv}_p < \epsilon} \quad  L\left( \x+\adv, y, \param \right) \right]
\end{equation}
The \textit{inner maximization} task is addressed by the attack algorithm utilized during adversarial training, and the \textit{outer minimization} is the standard ERM (\equationautorefname~\eqref{eq:cross-entropy}) employed to train the model parameterized with $\param$.
We separately apply both one-step FGSM (\equationautorefname~\eqref{eq:fgsm}), and $T$-step PGD (\equationautorefname~\eqref{eq:pgd}) algorithms to solve the inner maximization problem.
Throughout the remaining text, we refer to these as ``FGSM adversarial training'' and ``PGD-$T$ adversarial training'' respectively.

Notably, the overall training is done on clean as well as adversarial samples. The overall loss function is given by:
\begin{equation}
    \label{eq:overall_adv_train}
    L_\text{AT}(\x, \noisyx, y,\param ) = \left(1-w_{\text{AT}}\right) \cdot L(\x, y, \param) + w_{\text{AT}} \cdot L(\noisyx, y, \param),
\end{equation}
where $w_{\text{AT}}$ is the weight of the adversarial training.

% Second,
% we tested a similar but stronger scheme in which FGSM is replaced by PGD.
% In the PGD-10 we tested, the adversarial samples were generated from \eqref{eq:fgsm} with $i=10$. 
% namely, by running \eqref{eq:fgsm} for 10 times.

\paragraph{Adversarial Lipschitz Regularization (ALR).}
This approach of gaining robustness is based on learning a function that is not much sensitive to a small change in the input.
In other words, if we can learn a relatively smooth function, then the posterior distribution should not vary abruptly if the input perturbation is within the maximum allowed limit. 
We propose a training strategy equipped with the recently invented adversarial Lipschitz regularization technique \cite{DBLP:conf/iclr/Terjek20}.
Similar to the regularization based on local distribution smoothness in Virtual Adversarial Training (VAT) \cite{miyato2015distributional}, ALR imposes a regularization term defined using Lipschitz smoothness:
\begin{equation}
    \norm{f}_L = \sup_{\x, \noisyx ~\in X, d_X (\x, \noisyx)>0} 
        \frac{
            d_Y (f(\x), f(\noisyx) ) 
        }{
            d_X (\x, \noisyx)
        },
\end{equation}
where $f(\cdot)$ the function of interest (implemented by the neural network) that maps the input metric space $(X, d_X)$ to output metric space $(Y, d_Y)$.
In our case of speaker classification,
we chose $f(\cdot)$ as the final log-posterior output of the network, \ie $f(\x) = \log p(y|\x, \param)$,
$l_1$ norm as $d_Y$, and
$l_2$ norm as $d_X$.
The adversarial perturbation $\adv = \epsilon ~\adv_k$ in $\noisyx = \x + \adv$
is approximated by the power iterations:
\begin{equation}
% \begin{split}
    %  \adv       = \epsilon \adv_k, %\\
    ~\adv_{i+1} = \frac{
           \nabla_{\adv_i} d_Y \big( f(\x), f( \x + \xi \adv_i ) \big)
    }{
         \norm{\nabla_{\adv_i} d_Y \big( f(\x), f( \x + \xi \adv_i ) \big)}_2
    },
% \end{split}
\end{equation}
where, $\adv_0$ is randomly initialized, and $\xi$ is another hyperparameter (see Section~\ref{subsec:Defense parameters}).
The regularization term added to training is
\begin{equation}
\label{eq:alr}
    L_{ALR} = \left[ 
        \frac{
            d_Y (f(\x), f(\noisyx) ) 
        }{
            d_X (\x, \noisyx)
        }
        - K
    \right]_{+},
\end{equation}
where $K$ is the desired Lipschitz constant we wish to impose.

\section{Experimental Setting}\label{sec:Experimental setting}
We implement the core of most of our attack and defense algorithms (except ALR) through the Adversarial Robustness Toolbox~\cite{art2018}. For ALR, we follow the original implementation of~\cite{DBLP:conf/iclr/Terjek20}. The rest of the experimental details are described below.

\subsection{Dataset}
We employ Librispeech~\cite{panayotov2015librispeech} (the ``train-clean-100'' subset) dataset for all the experiments.
It contains $100$ hours of clean speech from $251$ unique speakers ($125$ females).
We utilize all the speakers for our experiment.
For every speaker, we employ $90\%$ of the utterances for training the classifier, and the remaining $10\%$ utterances for testing.
The train-test split is deterministic, and it is kept fixed throughout all the experiments.

\subsection{Model architectures}\label{subsec:Model architectures}
We implemented our classifier, $f: \X \rightarrow \Y$, by combining a Convolutional Neural Network (CNN) with a digital signal processing (DSP) front-end which is differentiable.
The \textit{non-trainable} DSP front-end extracts log Mel-spectrogram which is viewed as a temporal signal of $F$ channels, where $F$ is the number of Mel frequency bins.
The back-end is either of the two DNN models described below.
As both modules are differentiable, the adversarial attack schemes introduced in Section~\ref{sec:Methodology} can be applied to create time-domain perturbation directly.

\subsubsection{1D CNN} 
The model consists of 8 stacks of convolutional layers that transforms the Mel-spectrogram into the label space $\Y$.
ReLU nonlinearity and batch normalization are used after every convolutional layer.
Maxpool is employed after every alternate layer.
The penultimate layer has a dimension of 32.
The model has total of 219k trainable parameters.
We perform most of our analysis with this model, and utilize the following TDNN model for transferability experiments.

\subsubsection{TDNN} 
The Time Delay Neural Network~(TDNN)~\cite{snyder2017deep,snyder2018x} is one of the current state-of-the-art models for speaker recognition. 
We adopt the model architecture proposed in~\cite{snyder2018x} for the experiments related to transferability analysis (Section~\ref{subsec:Ablation study 3: Transferability analysis}).
The model consists of time-dilated convolutional layers along with a statistics pooling module, and it has $\sim 4.4$ million trainable parameters, and hence, is much larger than the 1D CNN model.

\subsection{Training parameters}
We employ the Adam optimizer~\cite{kingma2014adam} with a learning rate of $0.001$, $\beta_1=0.5$, and $\beta_2=0.999$.
We train with a minibatch size of $128$, and train all models (except ALR) for $30$ epochs, since the training accuracy reaches almost $100\%$ and saturates within that.
The ALR training converges much slowly, and thus, all ALR-based experiments are run for $2500$ epochs.
The training accuracy tends to saturate after that.

\subsection{Attack parameters}
Our main results (Section~\ref{subsec:Main result}) are obtained from the experiment with attack strength $\epsilon=0.002$ for $l_\infty$ attacks, and confidence margin $\delta=0$ for Carlini $l_2$ attack.
The choice of $\epsilon=0.002$ is due a reasonably strong SNR ($\sim 30dB$ for FGSM/PGD) of the adversarial samples (see Section~\ref{subsec:Attack strength vs. SNR}).
Furthermore, we vary the strength of different attacks, and the results are shown in Section~\ref{subsec:Ablation study 1: Varying attack strength}.
The PGD attack is for $100$ iterations with a step size $\alpha = \epsilon/5$.

% \subsection{Time-domain attack}
% Most of the state-of-the-art speaker recognition models are trained with mel-spectrograms~\cite{snyder2018x,Nagrani2017,Chung2018}, but employing mel-spectrograms in adversarial sample generation leads to generating perturbed samples in mel-spectrogram space (and not in time-domain signal space).
% This motivates us to generate mel-spectrogram online during training, and generate the adversarial samples in the time domain.
% In other words, the mel-spectrogram extraction is a deterministic mapping without any learnable parameters, and the gradient is computed with respect to the time domain signal during adversarial sample generation.
% Formally,
% \begin{align}
%     \boldsymbol{X} &= \text{\textmyfont{mel-spectrogram}}(\boldsymbol{x}) \\
%     \boldsymbol{\widetilde{x}} &= \text{\textmyfont{attack}}\left( \nabla_{\boldsymbol{x}} \left[ L\left( \boldsymbol{X}, y; \boldsymbol{\theta} \right) \right] \right)
% \end{align}

\subsection{Defense parameters}\label{subsec:Defense parameters}
In ALR method, we set the number of power iterations $K=1$, and the hyperparameter $\xi=10$, as recommended in~\cite{DBLP:conf/iclr/Terjek20}.
The FGSM- and PGD-based adversarial training algorithms are run with $\epsilon=0.002$.
Hence, the main results (Section~\ref{subsec:Main result}) employ the same $\epsilon$ value in both the attack and the adversarial training based defense.
The ablation study in Section~\ref{subsec:Ablation study 1: Varying attack strength} is particularly designed to investigate the effect of using different values of $\epsilon$ during the attack.
Specifically, the study varies $\epsilon$ above and below the vicinity of $\epsilon=0.002$ (set during defense training), and analyzes the effectiveness of the defense method.
The PGD adversarial training uses $10$ iterations (\ie PGD-10 as introduced in Section~\ref{subsec:Defense algorithms})\footnote{PGD adversarial training is slow, and we could not afford to run it for more than $10$ iterations.}, although we evaluated it against PGD attack with higher number of iterations (Section~\ref{subsec:Main result} and \ref{subsec:Ablation study 2: Analyzing the best defense method}). 
During adversarial training, we create minibatches containing equal number of clean and adversarial samples, \ie in \equationautorefname~\eqref{eq:overall_adv_train}, we set $w_\text{AT}=0.5$. 

\section{Results and Discussion}\label{sec:Results and discussions}

\subsection{Attack strength \textit{vs.} SNR} \label{subsec:Attack strength vs. SNR}
% Old -- too detailed, confusing
% \begin{figure}
%     \centering
%     \begin{subfigure}[b]{0.49\textwidth}
%         \centering
%         \includegraphics[width=\textwidth]{figs/snrs/snr_fgsm-pgd100-carliniInf_attack.pdf}
%         \caption{FGSM, PGD, and Carlini $l_\infty attacks$}   
%         \label{fig:SNR_fgsm_pgd_carliniInf_attack}
%     \end{subfigure}
%     \begin{subfigure}[b]{0.49\textwidth}  
%         \centering 
%         \includegraphics[width=\textwidth]{figs/snrs/snr_carlini_L2_attack.pdf}
%         \caption{Carlini $l_2$ attack}    
%         \label{fig:SNR_carlini_l2_attack}
%     \end{subfigure}
%     \caption{Mean SNR (standard deviation shown as band around the mean line) of the test adversarial samples generated by different attack algorithms.} 
%     \label{fig:SNR}
% \end{figure}

\begin{figure}
    \centering
    \begin{subfigure}[b]{0.49\textwidth}
        \centering
        \includegraphics[width=\textwidth]{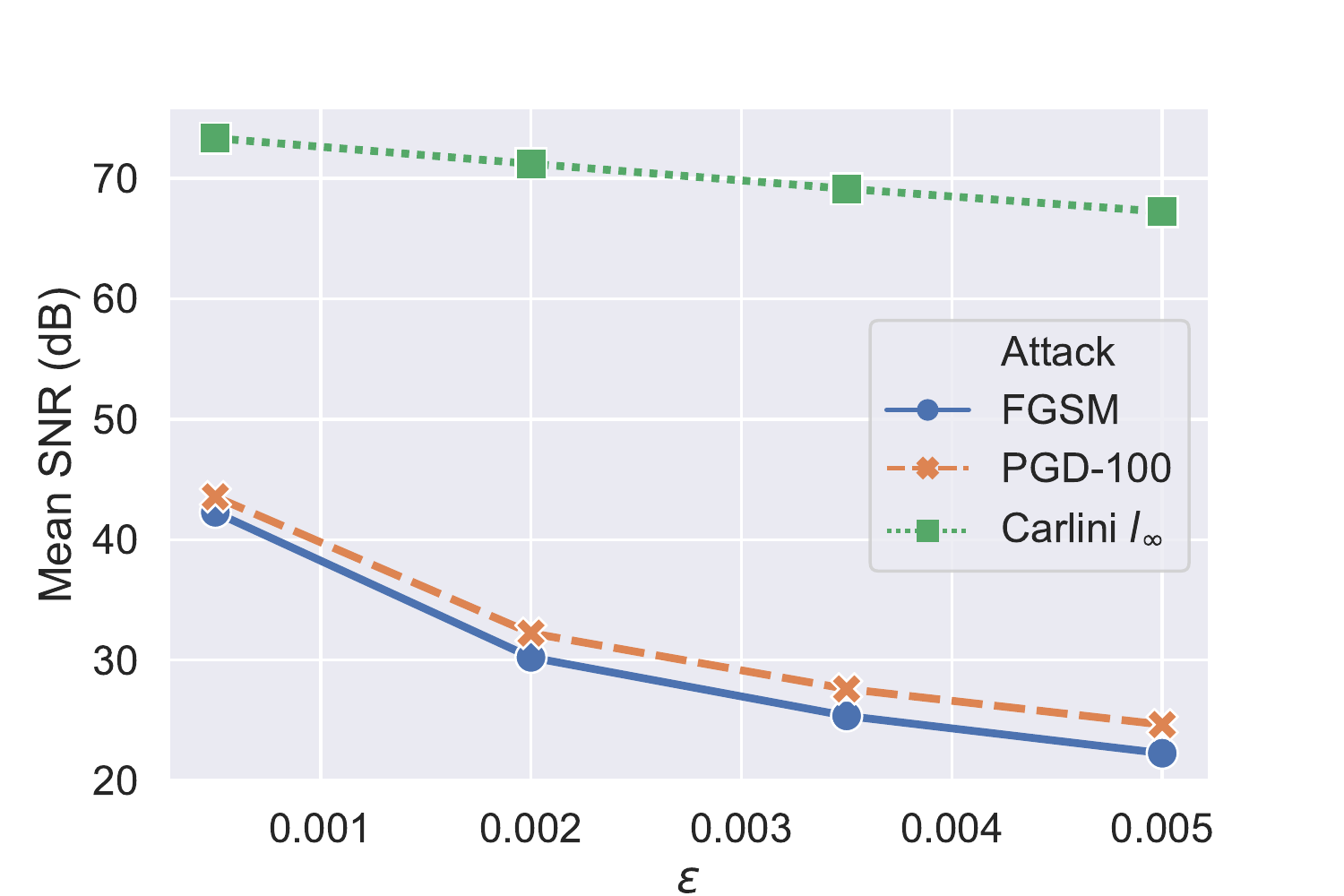}
        \caption{Attack strength \textit{vs.} SNR}   
        \label{fig:SNR}
    \end{subfigure}
    \begin{subfigure}[b]{0.49\textwidth}  
        \centering 
        \includegraphics[width=\textwidth]{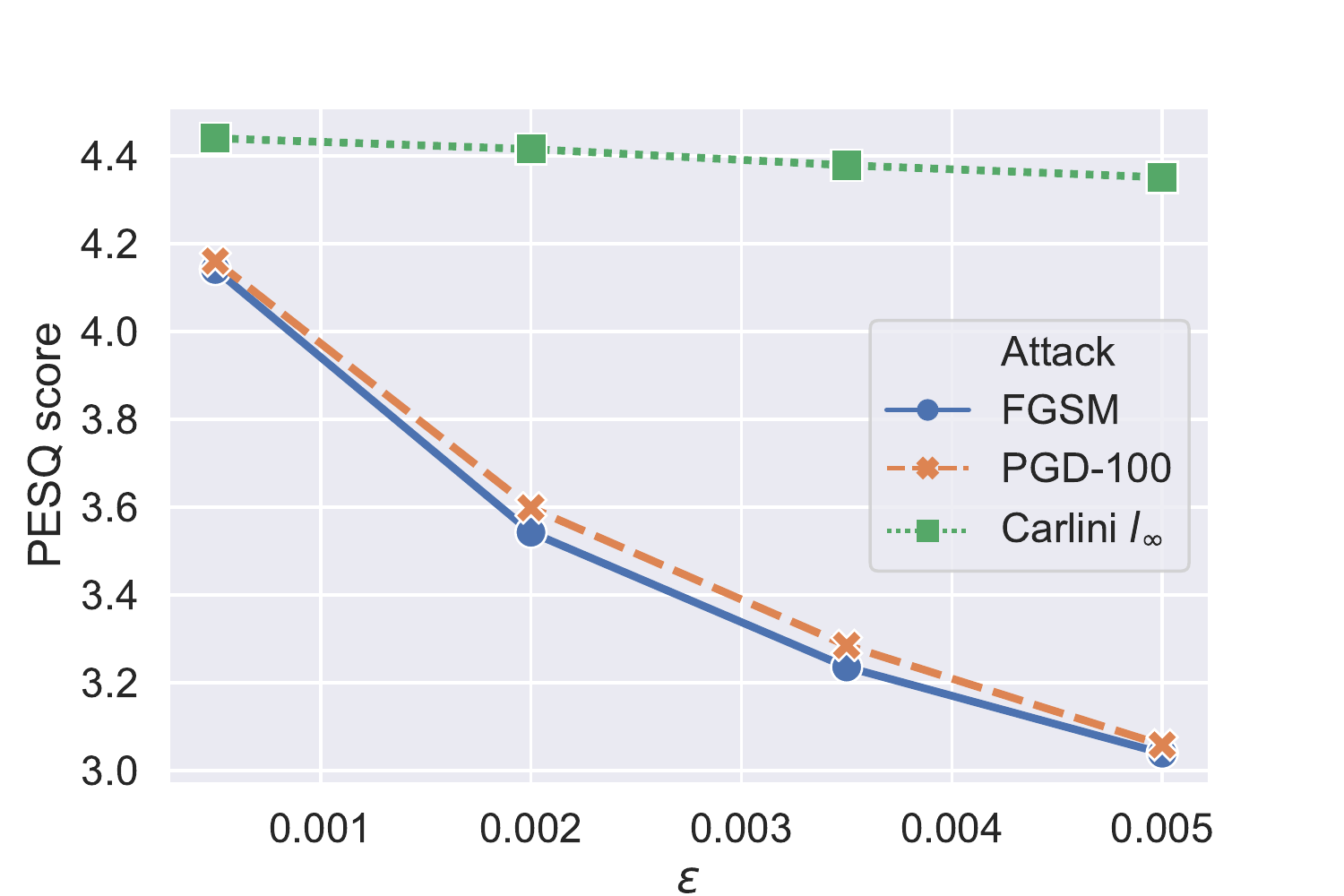}
        \caption{Attack strength \textit{vs.} perceptibility}    
        \label{fig:PESQ}
    \end{subfigure}
    \caption{Mean SNR and PESQ score of the test adversarial samples generated by different $l_\infty$ attack algorithms at different strengths.} 
    \label{fig:SNR and PESQ}
\end{figure}

To have a substantial understanding about the strength of different attack algorithms, we computed the mean Signal-to-Noise Ratio (SNR) of all the test adversarial samples for every level of attack strength.
For $l_\infty$ attacks, $\epsilon$ varies between $\{0.0005, 0.002, 0.0035, 0.005\}$, and for Carlini $l_2$ attack the confidence margin $\delta$ varies between $\{0, 0.001, 0.01, 0.1\}$.
The curves for $l_\infty$ attacks are shown in \figureautorefname~\ref{fig:SNR}.
%explanation for the old, detailed figure -- abandoned
%%FGSM and PGD attacks generate adversarial samples with mean SNR approximately equal for all models, but Carlini $l_\infty$ and $l_2$ attacks generate perturbed samples at different mean SNRs for differently trained models (\ie differently defended models).
%%We can see that the standard deviation bands are extremely narrow (not visible) for FGSM and PGD attacks, but those bands are much wider for Carlini methods.
There are two important observations.
First, the average level of SNR is $\sim 30$ dB higher for Carlini $l_\infty$ than FGSM and PGD.
Second, the SNR level tends to decrease faster with increase in $\epsilon$ for PGD and FGSM as compared to Carlini $l_\infty$.
The reason might be attributed to the optimization algorithms that various attack methods use for generating the adversarial samples. For example, the Carlini method enforces minimum perturbation required to change the output prediction, while PGD enforces perturbation projected inside the $l_\infty$-ball of size $\epsilon$ around $\x$.
We have also computed mean SNR for Carlini $l_2$ attack for different values of the confidence margin $\delta$. The SNR level tends to stay around $75$ dB, and does not vary much with increasing $\delta$.
A visualization of the adversarial spectrograms is shown in \appendixautorefname~\ref{app:Visualizing spectrograms} for a more detailed analysis of the attack algorithms.

\subsection{Attack strength \textit{vs.} perceptibility} \label{subsec:Attack strength vs. perceptibility}
We measure the perceptibility of the generated adversarial samples by employing Perceptual Evaluation of Speech Quality (PESQ)~\cite{recommendation2001perceptual,rix2001perceptual}.
While subjective measure with multiple human annotators can be more accurate, it is time-consuming and costly.
The objective PESQ measure has been the ITU-T standard for measuring telephonic transmission quality.
It gives a mean opinion score by comparing the degraded speech signal with the original speech recording.
The PESQ score is between $-0.5$ to $4.5$, and a higher value indicates better quality.
Figure~\ref{fig:PESQ} depicts the average PESQ scores of all the test adversarial samples generated via different attack methods at various strengths.
We can see the gradual degradation of audio quality with the increase of attack strength.
Similar to the findings in the SNR analysis (Section~\ref{subsec:Attack strength vs. SNR}), Carlini $l_\infty$ attack produces more perceptible adversarial audio samples than PGD-100 and FGSM.
The degradation is also slower for Carlini attack.
It is noteworthy that at $\epsilon=0.0005$ the attack algorithms are able to achieve high audio quality (PESQ score $> 4.0$), but force the classifier to produce erroneous outputs (Section~\ref{subsec:Ablation study 1: Varying attack strength}).
We have also computed the average PESQ score for Carlini $l_2$ attack.
The PESQ score is $\sim 4.4$, and does not vary much with change in the confidence margin $\delta$.

\subsection{Main results}\label{subsec:Main result}
Table~\ref{tab:main_results} presents the test performance of standard training (without any defense) and all the employed defense methods for three $l_\infty$ attacks, and one $l_2$ attack.
All the performances are averaged over $10$ random runs.
The $l_\infty$ attacks are with $\epsilon=0.002$, and all the adversarial training methods are run with the same $\epsilon$ value.
As we can see, the accuracy of the standard training method drops from $94\%$ by a significant margin for all the attacks.
This shows the vulnerability of the model, and further underscores the need for strong countermeasures.
A comparison between the three $l_\infty$ attacks shows the FGSM is the weakest one ($25\%$ adversarial accuracy for standard training), and PGD-100 (PGD with 100 iterations) is the strongest one ($0\%$ adversarial accuracy for standard training). 

\begin{table}
\caption{Different attacks on a speaker recognition system, and performance of different defense methods. ``\textit{Benign}'' denotes accuracy on clean samples, and ``\textit{Adv.}'' denotes accuracy on adversarial samples. Accuracy is on a scale of $[0,1]$.}
 \label{tab:main_results}
\centering
 \begin{tabular}{cc | cc |cc |cc |cc } 
   \toprule
   & & \multicolumn{2}{c|}{\makecell{\textbf{Standard} \\\textbf{training}}} & \multicolumn{2}{c|}{\makecell{\textbf{FGSM} \\\textbf{adv. training}}} & \multicolumn{2}{c|}{\textbf{ALR}} & \multicolumn{2}{c}{\makecell{\textbf{PGD-10} \\\textbf{adv. training}}}\\ \midrule
  \textbf{Norm} & \textbf{Attack} & \textit{Benign} & \textit{Adv.} & \textit{Benign} & \textit{Adv.} & \textit{Benign} & \textit{Adv.} & \textit{Benign} & \textit{Adv.} \\ \midrule
  \multirow{3}{*}{$l_\infty$} & FGSM & \multirow{4}{*}{0.94} & 0.25 & \multirow{4}{*}{0.82} & 0.20 & \multirow{4}{*}{\textbf{0.96}} & 0.44 & \multirow{4}{*}{0.92} & \textbf{0.73} \\
  & Carlini $l_\infty$ & & 0.02 & & 0.09 & &	0.10 & & \textbf{0.58} \\
  & PGD-100 & & 0.00 & & 0.00 & & 0.00 & & \textbf{0.43} \\\midrule
  $l_2$ & Carlini $l_2$ & & 0.00 & & 0.00 & & 0.00 & & \textbf{0.09} \\\bottomrule
 \end{tabular}
\end{table}

Comparing different defense methods, we can see that FGSM-based adversarial training is the weakest defense strategy.
The ALR method is better than FGSM adversarial training, although it fails to defend against a PGD-100 attack.
The PGD-10 adversarial training is found to perform the best in our experiments.
It is interesting to see that PGD adversarial training with $10$ iterations is able to defend against a PGD attack with $100$ iterations.
The proposed PGD-10 adversarial training gives absolute improvements of $48\%, 56\%$ and $43\%$ over the undefended performance against FGSM, Carlini $l_\infty$ and PGD-100 attacks, respectively.

As observed in the previous literature, the performance gain achieved by PGD-10 adversarial training against different adversarial attacks generally results in a drop in benign accuracy.
Similarly, in our experiment, the accuracy on the clean test samples drops for both FGSM- and PGD-based adversarial training methods, with the FGSM variant getting higher drop in performance.
The ALR method, on the other hand, achieves a $2\%$ absolute improvement in benign accuracy compared to the model with standard training, possibly because of lesser overfitting due to the presence of the penalty term shown in \equationautorefname~\eqref{eq:alr}.

The last row of Table~\ref{tab:main_results} shows the performance of different defense methods against Carlini $l_2$ attack. 
Standard training, FGSM adversarial training, and ALR algorithms are unable to defend against this attack.
Defense with PGD-10 adversarial training also performs poorly for this $l_2$ attack. 
The reason might be attributed to the adversarial training methodology which is based on $l_\infty$ perturbation, and thus, probably fails to defend against a strong $l_2$ attack.

A related ablation study is provided in \appendixautorefname~\ref{app:Similarity in misclassification for different attacks} which shows the similarity between the misclassified predictions made by the model under different attacks. This could possibly reveal some inherent similarities between different attacks.

\begin{figure*}
    \centering
    \begin{subfigure}[b]{0.49\textwidth}
        \centering
        \includegraphics[width=\textwidth]{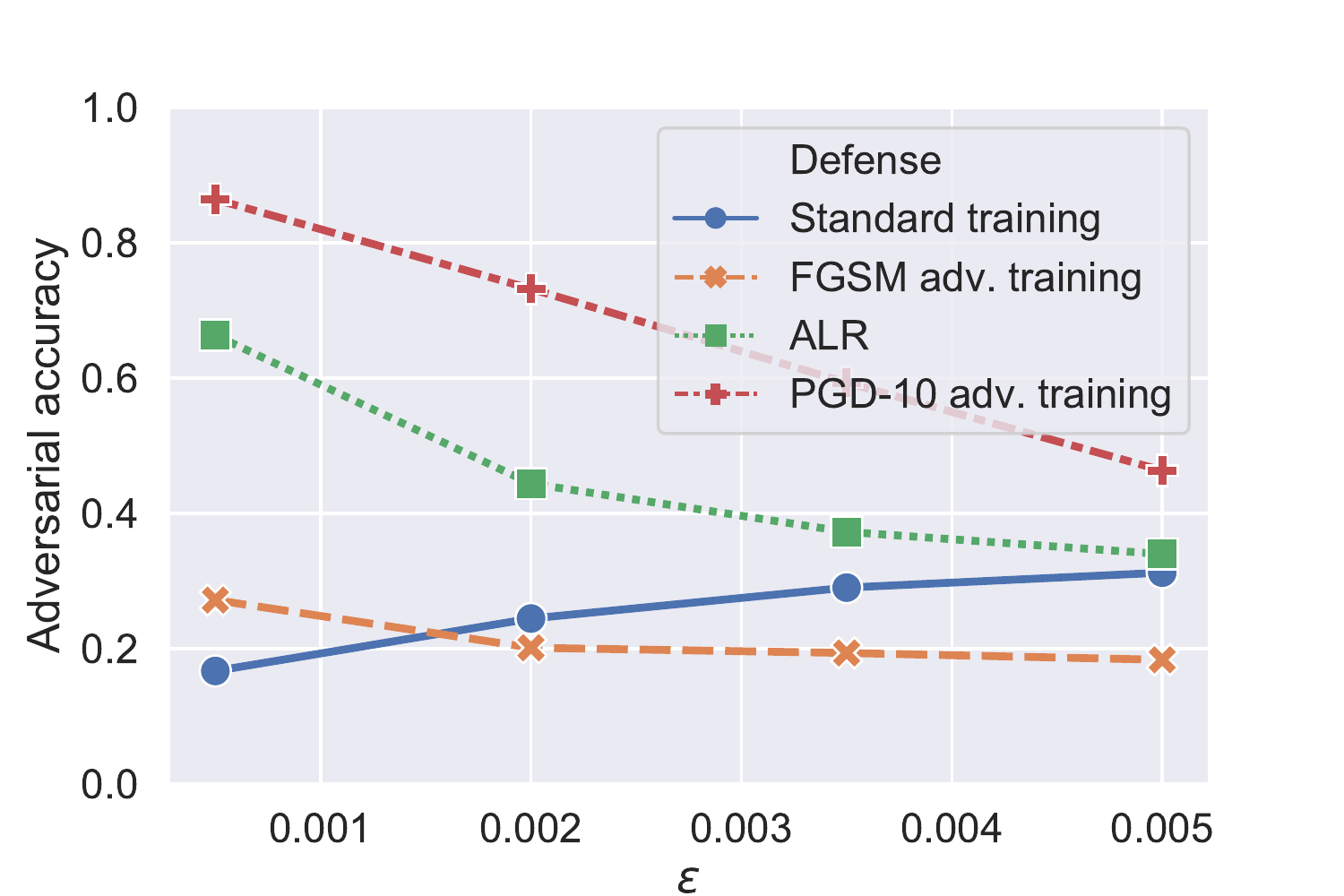}
        \caption{FGSM attack}   
        \label{fig:fgsm_attack}
    \end{subfigure}
    \begin{subfigure}[b]{0.49\textwidth}  
        \centering 
        \includegraphics[width=\textwidth]{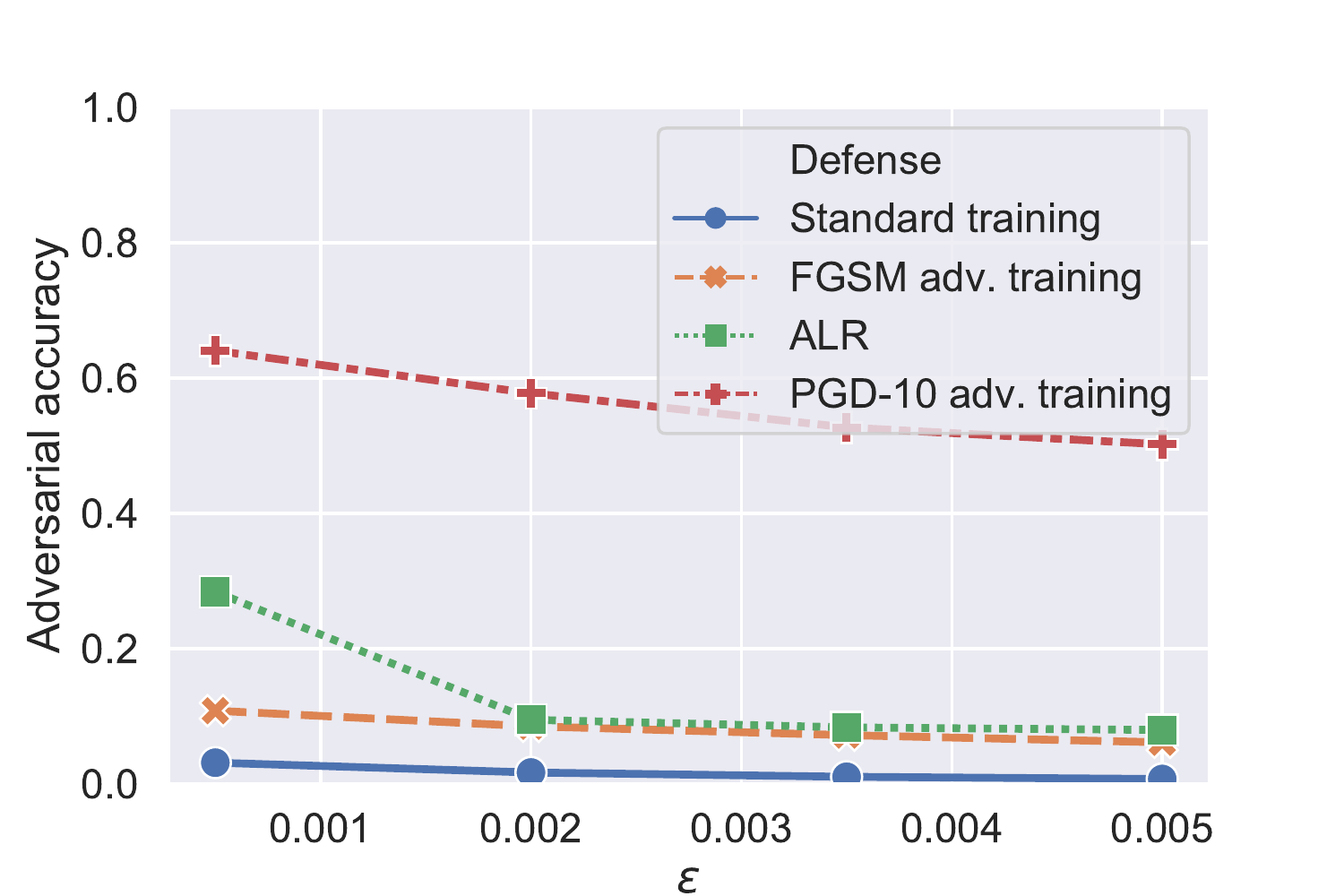}
        \caption{Carlini $l_\infty$ attack}    
        \label{fig:carlini_linf_attack}
    \end{subfigure}
    \\
    \begin{subfigure}[b]{0.49\textwidth}   
        \centering 
        \includegraphics[width=\textwidth]{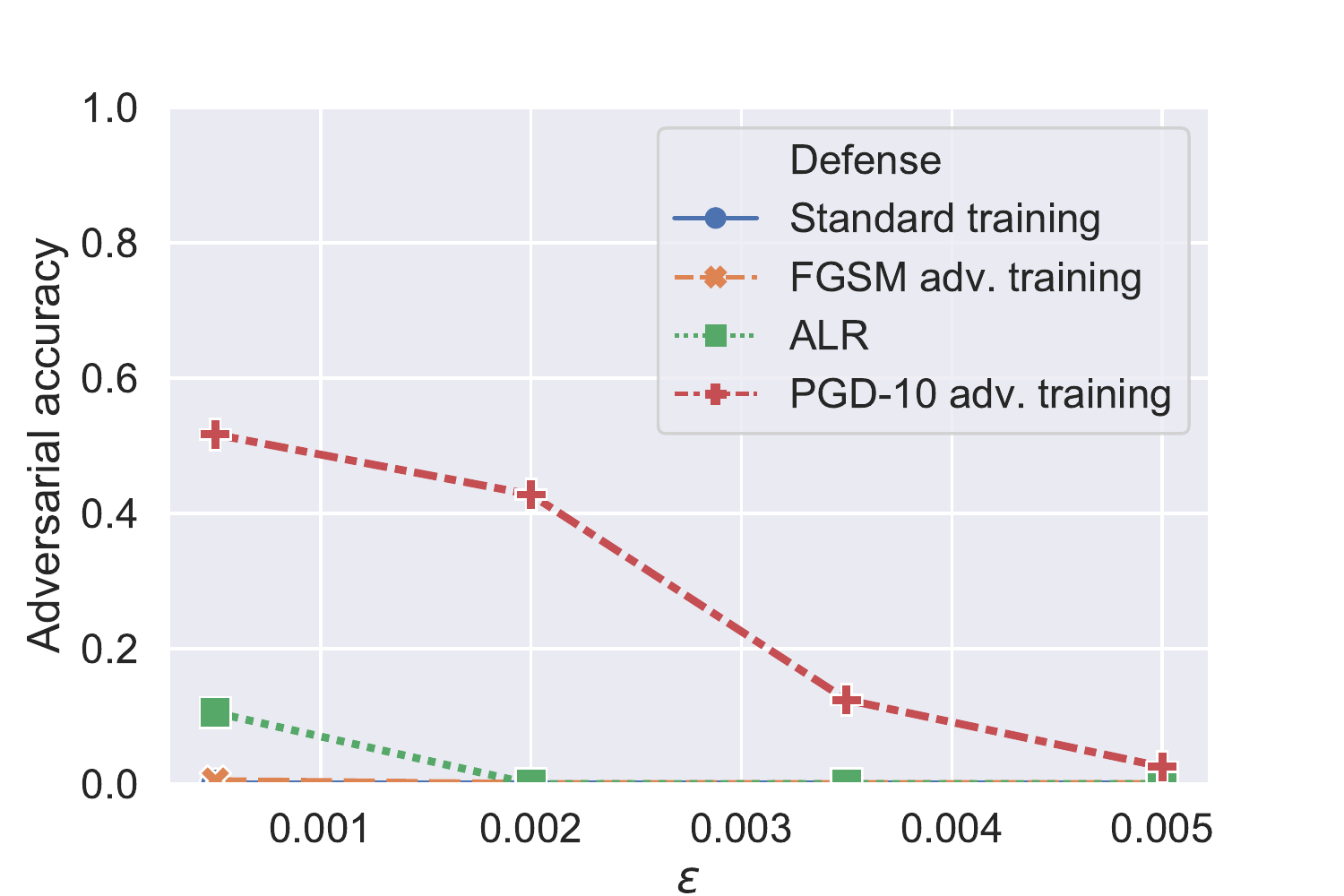}
        \caption{PGD-100 attack}    
        \label{fig:pgd100_attack}
    \end{subfigure}
    \begin{subfigure}[b]{0.49\textwidth}   
        \centering 
        \includegraphics[width=\textwidth]{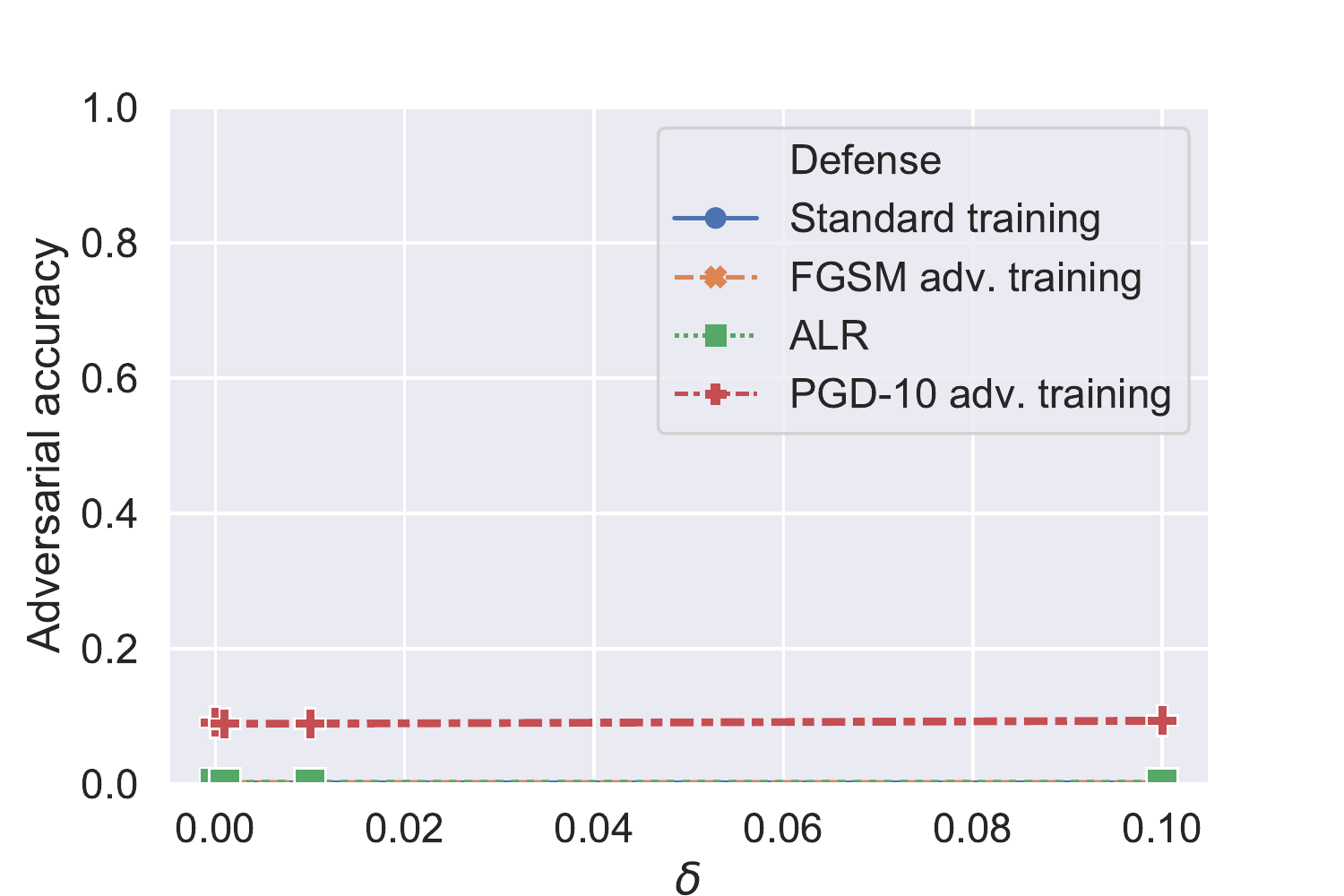}
        \caption{Carlini $l_2$ attack}    
        \label{fig:carlini_l2_attack}
    \end{subfigure}
    \caption{Ablation study 1: Varying the strength ($\epsilon$ for three $l_\infty$ attacks, $\delta$ for the Carlini $l_2$ attack) in different attack algorithms, and performance of different defense methods.} 
    \label{fig:ablation_study1}
\end{figure*}
\subsection{Ablation study 1: Varying attack strength}\label{subsec:Ablation study 1: Varying attack strength}
Figure~\ref{fig:ablation_study1} shows how the performances of different defense methods vary when we vary the strength of the adversarial attacks.
Note that the adversarial training-based defense methods still employ the same $\epsilon=0.002$ during training, but $\epsilon$ of the attack algorithm varies.
%%For $l_\infty$ attacks, $\epsilon$ varies between $\{0.0005, 0.002, 0.0035, 0.005\}$, and for Carlini $l_2$ attack the confidence margin $\delta$ varies between $\{0, 0.001, 0.01, 0.1\}$.

We can observe that the general trend of the curves is downward with the increase of the strength of any attack.
The only exception is the standard training scenario for FGSM attack. The performance surprisingly increases in the beginning and then saturates. 
%Moreover, in our experiment, FGSM-based adversarial training is not found to be effective.

Comparing different defense methods, we can see the PGD-10 adversarial training continues to outperform all the other defense methods for all attack types, and for all strength levels.
The proposed ALR training is found to be the next best defense technique.

Another interesting observation is that the accuracy curves for both the Carlini methods are more flat in nature compared to FGSM and PGD attacks. 
The reason might be attributed to the relatively less drop in SNR values of the test adversarial samples generated by Carlni method as the attack strength increases, as explained in Section~\ref{subsec:Attack strength vs. SNR}.

\subsection{Ablation study 2: Analyzing the best defense method}\label{subsec:Ablation study 2: Analyzing the best defense method}
\begin{figure}
    \centering
    \includegraphics[width=0.5\textwidth]{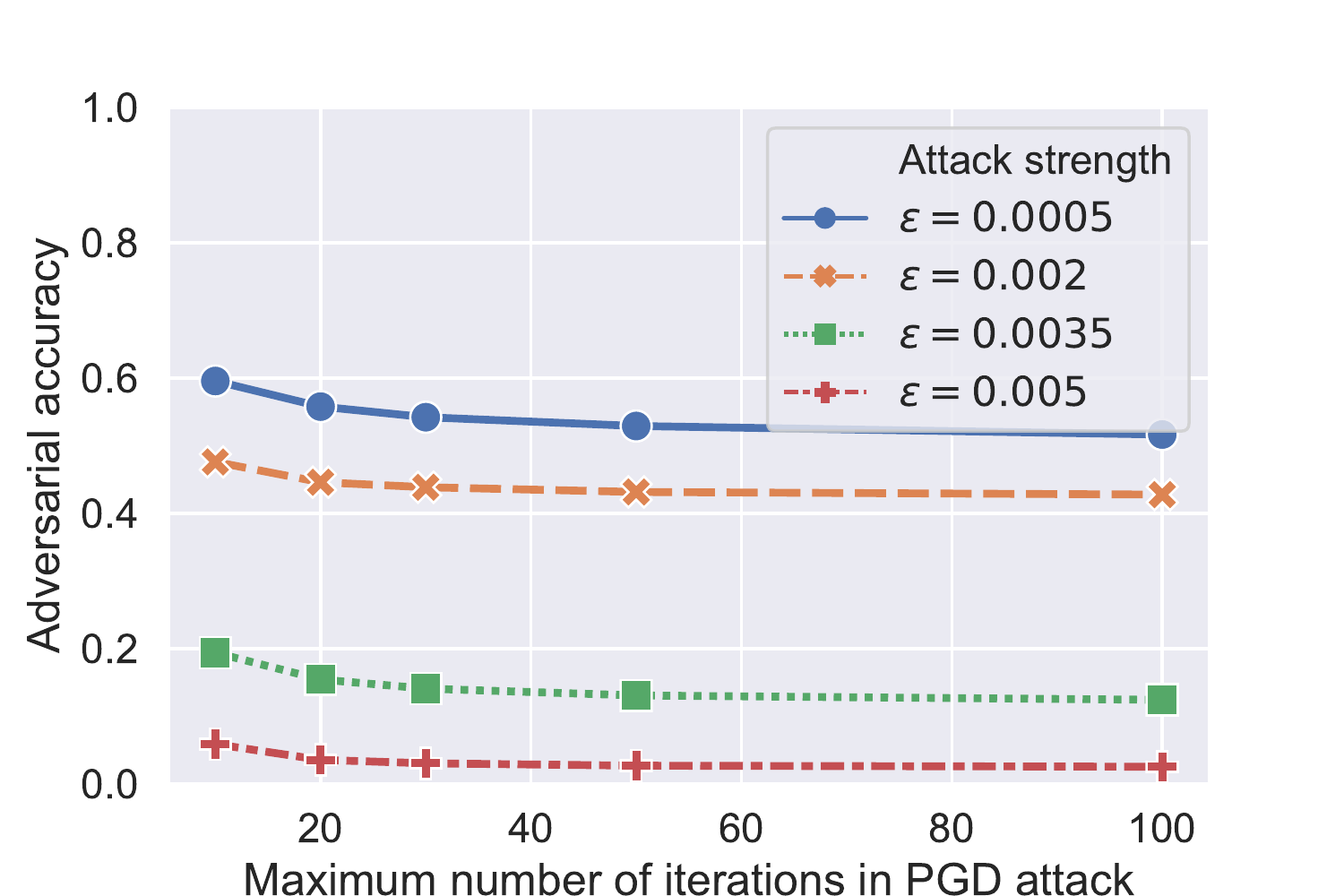}
    \caption{Performance of PGD-10 adversarial training against PGD attack at different strengths, and with different number of iterations.}
    \label{fig:pgd_detail}
\end{figure}
Here we analyze the best defense method, \ie PGD-10 adversarial training, in further detail.
Specifically, we investigate its behavior when we attack it with PGD attack with different number of iterations and at different strengths.
Figure~\ref{fig:pgd_detail} shows the variation of the adversarial accuracy for PGD-10 defense.
Each line denotes attack at a particular strength \ie a particular $\epsilon$ value.
The horizontal axis denotes the iteration number, varying in the range $\{10, 20, 30, 50, 100\}$.
A closer inspection reveals that after the first drop in performance for PGD-10 attack to PGD-20 attack, the accuracy value tends to decrease very slowly.
For example, at $\epsilon=0.002$, adversarial accuracy against PGD-10 attack is $\sim 48\%$.
Then a $3\%$ absolute drop is observed when we perform a PGD-20 attack, and the adversarial accuracy becomes $\sim 45\%$.
The accuracy tends to drop very slowly afterwards, and we see a $\sim 43\%$ accuracy against a PGD-100 attack.
We hypothesize that this behavior happens because around $20$ to $30$ iterations might be enough to project the perturbed adversarial samples to the edge of the $l_\infty$-ball, and thus much higher number of iterations do not necessarily produce a stronger attack.

\subsection{Ablation study 3: Transferability analysis}\label{subsec:Ablation study 3: Transferability analysis}

\begin{table}
\caption{Transferability of adversarial samples between different models. The adversarial samples are generated with the ``source'' model, but seem to be effective against the ``target'' model as well. Accuracy is on a scale of $[0,1]$.}
\label{tab:transferability}
\centering
    \begin{tabular}{cc | cc | cc | cc} 
        \toprule
        
        %& & & & \multicolumn{2}{c|}{\makecell{Source = 1D CNN,\\Target = TDNN}} & \multicolumn{2}{c}{\makecell{Source = TDNN,\\Target = 1D CNN}} \\ \toprule
        
        & & \multicolumn{2}{c|}{\textbf{Benign accuracy}} & \multicolumn{2}{c|}{\textbf{Adversarial accuracy}} & \multicolumn{2}{c}{\textbf{Adversarial accuracy}} \\ \midrule 
        
        \textbf{Source attack} & $\boldsymbol{\epsilon}$ & \textit{1D CNN} & \textit{TDNN} & \makecell{\textbf{Source}\\\textit{1D CNN}} & \makecell{\textbf{Target}\\\textit{TDNN}} & \makecell{\textbf{Source}\\\textit{TDNN}} & \makecell{\textbf{Target}\\\textit{1D CNN}} \\ \midrule
    
        FGSM & \multirow{2}{*}{0.002} & \multirow{2}{*}{0.98} & \multirow{2}{*}{0.87} & 0.19 & 0.40 & 0.03 & 0.16\\
        
        PGD-100	& & & &	0 & 0.36 & 0 & 0.08\\
        
        \bottomrule
        
    \end{tabular}
\end{table}

We perform the transferability analysis between the smaller CNN model, and the larger TDNN model (please see Section~\ref{subsec:Model architectures} for model architectures).
Table~\ref{tab:transferability} shows the benign accuracies for both the models\footnote{The TDNN model has lower benign accuracy possibly because of overfitting, but we do not spend time to fine-tune the TDNN model. Transferability of the adversarial samples is still clear from the table.}.
We can see that adversarial samples crafted from the ``source'' model tend to be harmful for the ``target'' model as well, as evident from the significant drop in the performances.
Adversarial samples generated with the larger model (TDNN) tend to be more effective in attacking the smaller model. 
Further studies are needed to fully understand the observed pattern of transferability.

\subsection{Ablation study 4: Effect of noise augmentation}
Noise augmentation is a standard technique employed during training a speaker recognition model~\cite{snyder2018x,jati2019multi}.
Here, we experiment with augmenting the dataset with white Gaussian noise (scaled with a factor equal to the $\epsilon$ used in the attack) during training the \textit{undefended} model.
The model is trained with both clean and noisy samples.
The experimental observations are tabulated in Table~\ref{tab:augmentation}.
As expected, the benign accuracy improves.
However, we could not find any improvement in the performance of the model for defending against FGSM and PGD-100 adversarial attacks.
The reason might be attributed to the ability of attack algorithms to generate more novel noise samples (compared to simple white Gaussian noise) that force the model posteriors to change.

\begin{table}
\caption{Effect of training data augmentation with white Gaussian noise. Accuracy is on a scale of $[0,1]$.}
\label{tab:augmentation}
\centering
    \begin{tabular}{cc | cc | cc} 
        \toprule
       
        & & \multicolumn{2}{c|}{\textbf{Benign accuracy}} & \multicolumn{2}{c}{\textbf{Adversarial accuracy}} \\ \midrule 
        
        \textbf{Attack} & $\boldsymbol{\epsilon}$ & \textit{No augmentation} & \textit{Augmentation} & \textit{No augmentation} & \textit{Augmentation} \\ \midrule
        
        FGSM & \multirow{2}{*}{0.002} & \multirow{2}{*}{0.94} & \multirow{2}{*}{\textbf{0.95}} & 0.25 & 0.17 \\
        PGD-100 & & & & 0 & 0\\
        
        \bottomrule
        
    \end{tabular}
\end{table}

\section{Conclusion and Future Directions}\label{sec:Conclusion and future directions}
The paper presented an extensive exploratory analysis of adversarial attacks on a closed set speaker recognition system. We reported results obtained from experiments with multiple state-of-the-art attack algorithms with varying attack strengths. 
We also investigated state-of-the-art defense methods, and adopted them for employing as countermeasures for the speaker recognition model.
We performed several ablation studies to understand the SNR characteristics and perceptibility of the adversarial speech, analyze the transferability of the adversarial attacks, and the effectiveness of white noise augmentation during training.
The main observations are the following:
\begin{itemize}
    \item Speaker recognition system such as the one employed in the current study is vulnerable to white box adversarial attacks. The performance of the undefended model dropped from $94\%$ to $0\%$ with the strongest attack (PGD-100) in our experiment even at $40$ dB SNR and PESQ score $>4$.
    \item Adversarial samples crafted with the Carlini and Wagner method are found to have the best perceptual quality in terms of the PESQ score.
    \item The adversarial samples generated with a particular source model are found to transfer well to a different target model, and hence, are also harmful for the target model. This is particularly alarming because it can open up chances for \textit{black box} attacks.
    \item Augmenting training data with white Gaussian noise is \textit{not} found to be effective.
    \item Experimenting with several defense methods showed that PGD-based adversarial training is the best defense strategy in out setting. 
    \item Although, PGD adversarial training is the best defense method, it is \textit{not} found to be effective against $l_2$ attack in our experiments, probably because of employing $l_\infty$ norm during training.
\end{itemize}
We hope the source codes published along with this paper can be helpful to the research community interested in pursuing further work in this domain. Several important future directions can be taken from here.
\begin{itemize}
    \item Extending the work for a speaker verification setting would a good exploratory direction. Specifically, an end-to-end system like~\cite{wang2019adversarial} can be investigated with the state-of-the-art attacks introduced in this paper, and performances of different defense algorithms can be established.
    \item Metric learning such as triplet training~\cite{schroff2015facenet} are shown to learn compact and robust embeddings against adversarial attacks for images~\cite{mao2019metric,zhong2019adversarial}. Metric learning is also found to be useful for learning robust speaker embeddings in~\cite{jati2019multi}. A natural extension can be to verify the adversarial robustness of speaker embeddings learning via metric learning.
    \item Adaptive attacks~\cite{tramer2020adaptive} are particularly designed to break any specific defense algorithm. The strategies introduced in~\cite{tramer2020adaptive} can be a starting point to perform model-specific adversarial attacks on existing defense methods proposed for speaker recognition systems.
    \item Studying targeted attacks might be another good direction from here, especially, since this could be a potential threat for biometric systems that rely on speaker recognition modules.
    \item Finally, further research can be done on crafting imperceptible (to human judgement or by retaining high PESQ score) adversarial audio samples with high attack success rate such as in~\cite{qin2019imperceptible}, and also formulating effective detection~\cite{speakman2018subset} and defense algorithms as countermeasures.
\end{itemize}

%\newpage
\bibliographystyle{unsrt}  
\bibliography{mybibfile}

\appendixpage
\begin{appendices}
\begin{figure*}[h]
    \centering
    \begin{subfigure}[b]{0.49\textwidth}
        \centering
        \includegraphics[width=\textwidth]{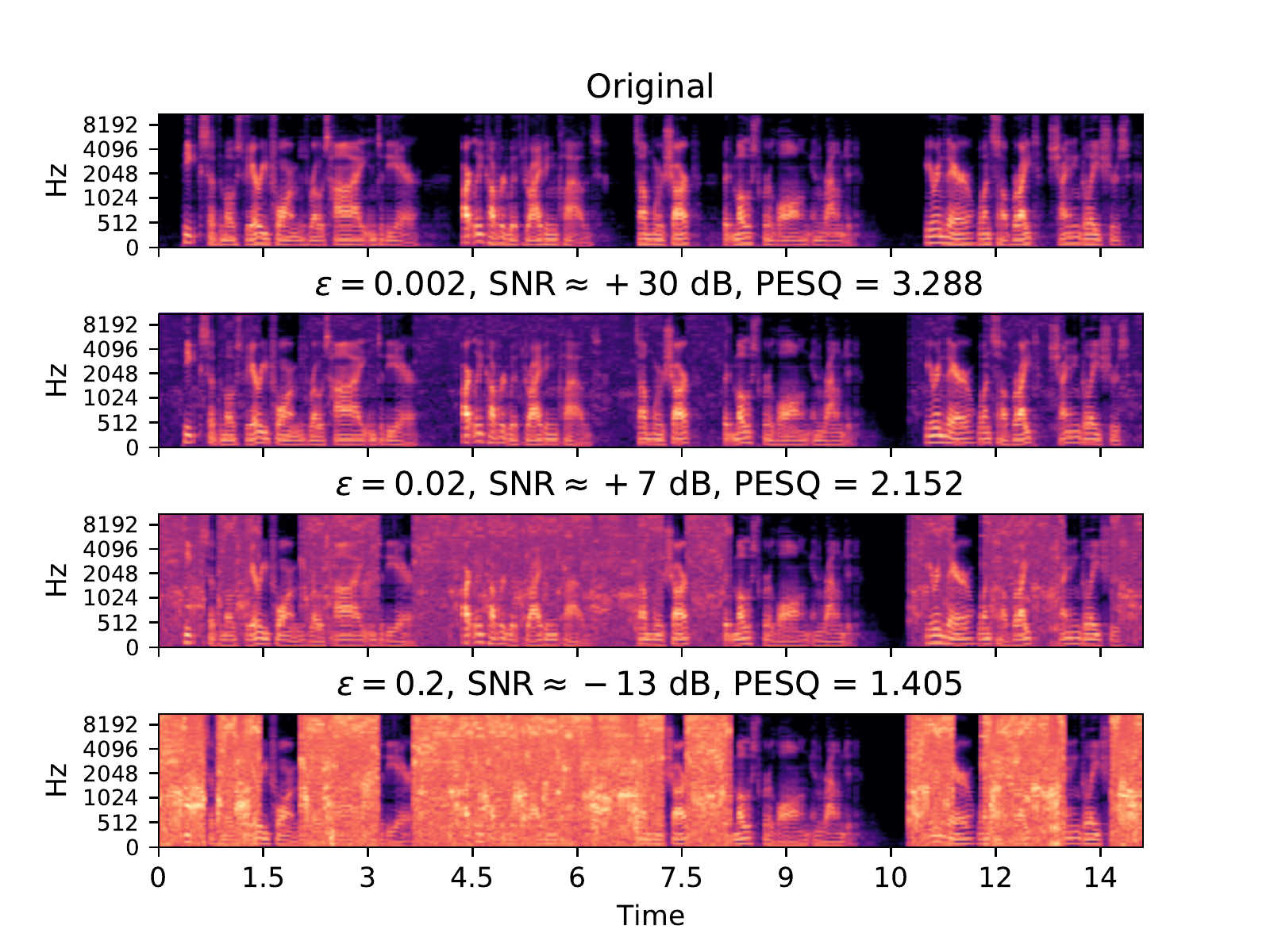}
        \caption{FGSM}   
        \label{fig:specs_fgsm}
    \end{subfigure}
    \begin{subfigure}[b]{0.49\textwidth}  
        \centering 
        \includegraphics[width=\textwidth]{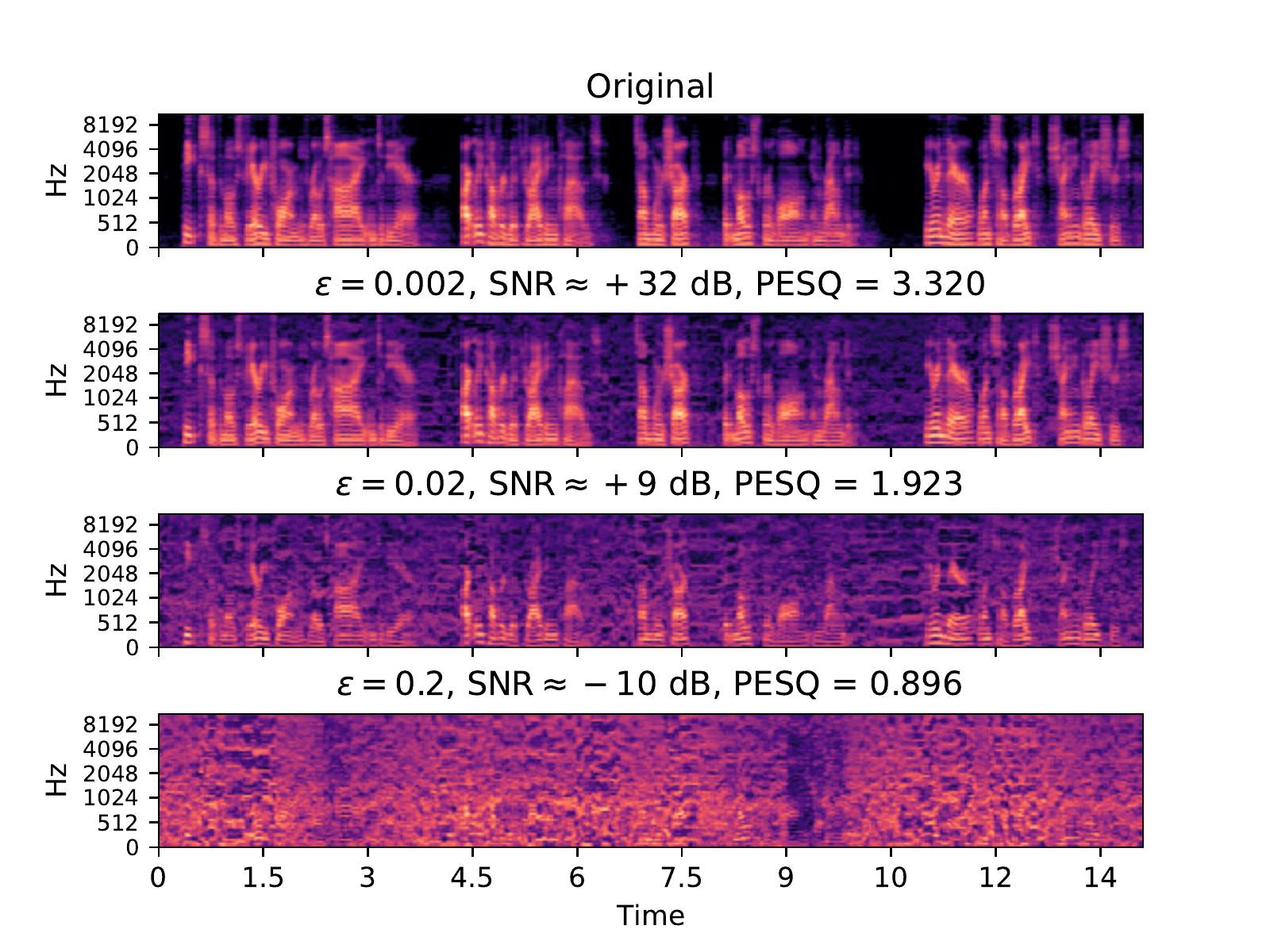}
        \caption{PGD-100}    
        \label{fig:specs_pgd}
    \end{subfigure}
    \\
    \begin{subfigure}[b]{0.49\textwidth}   
        \centering 
        \includegraphics[width=\textwidth]{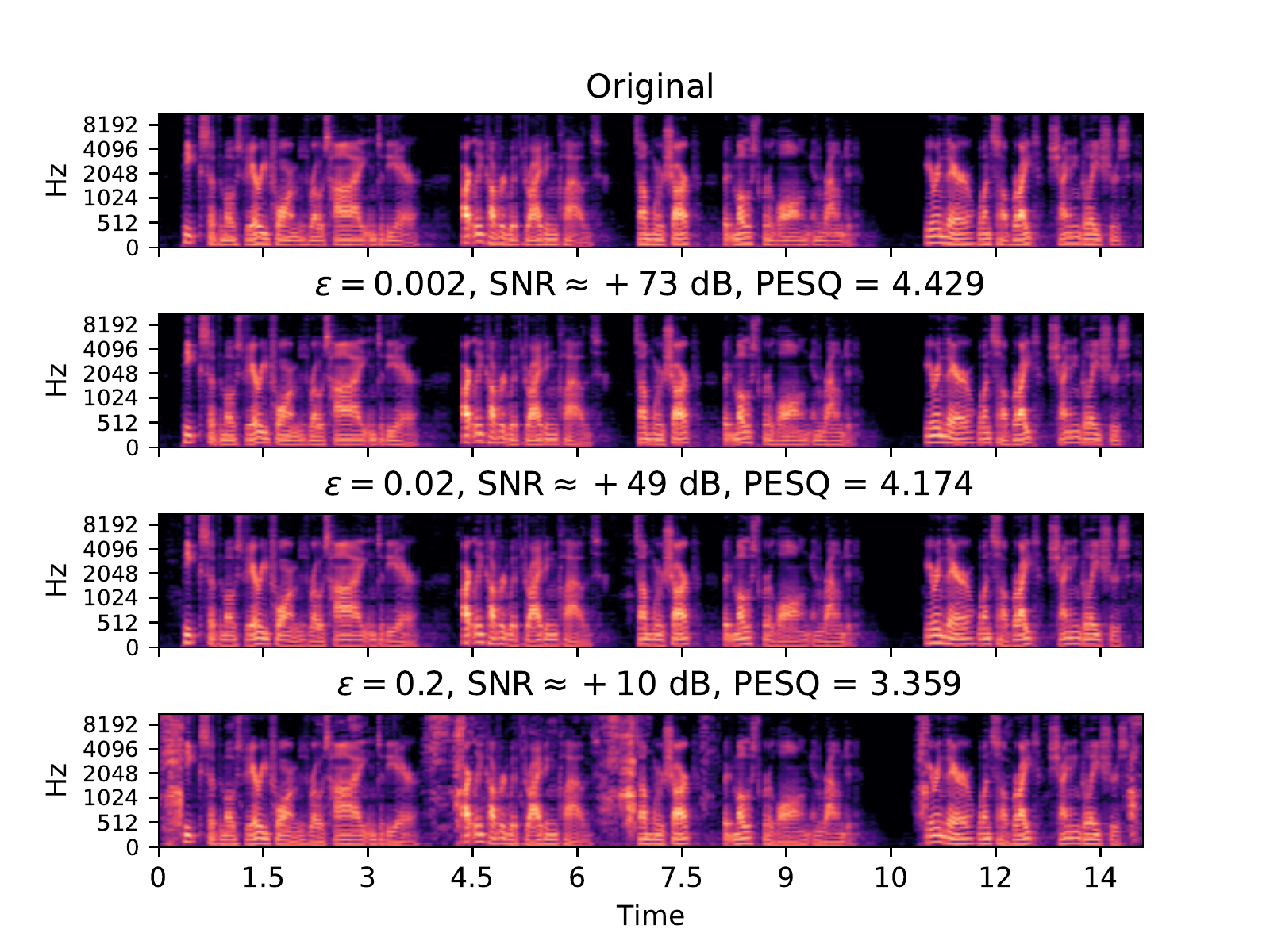}
        \caption{Carlini $l_\infty$}    
        \label{fig:specs_carliniLinf}
    \end{subfigure}
    \caption{Spectrograms of an original utterance and its perturbed versions under different $l_\infty$ attacks at varying strengths.} 
    \label{fig:app spectrograms}
\end{figure*}

\begin{figure*}[!h]
    \centering
    \begin{subfigure}[b]{0.33\textwidth}
        \centering
        \includegraphics[width=\textwidth]{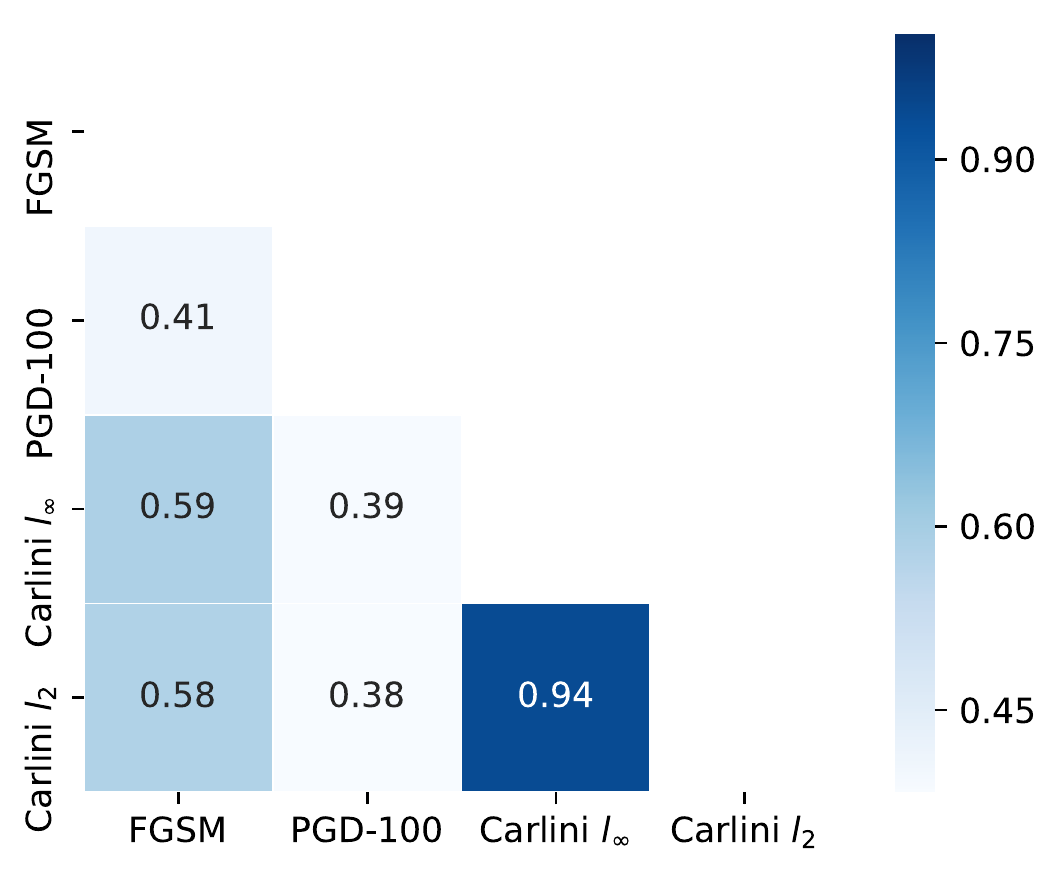}
        \caption{$\epsilon=0.0005$}   
        \label{fig:cmat_eps_0.0005}
    \end{subfigure}
    \begin{subfigure}[b]{0.33\textwidth}  
        \centering 
        \includegraphics[width=\textwidth]{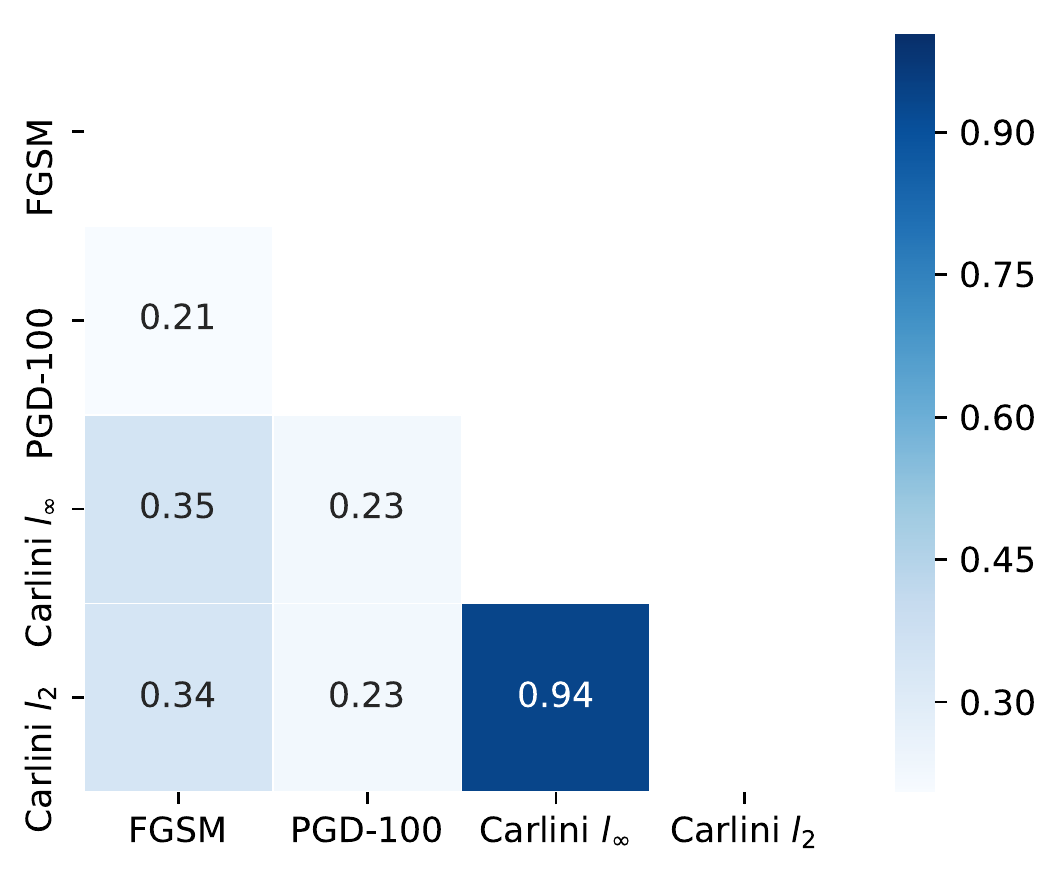}
        \caption{$\epsilon=0.002$}    
        \label{fig:cmat_eps_0.002}
    \end{subfigure}
    \begin{subfigure}[b]{0.33\textwidth}   
        \centering 
        \includegraphics[width=\textwidth]{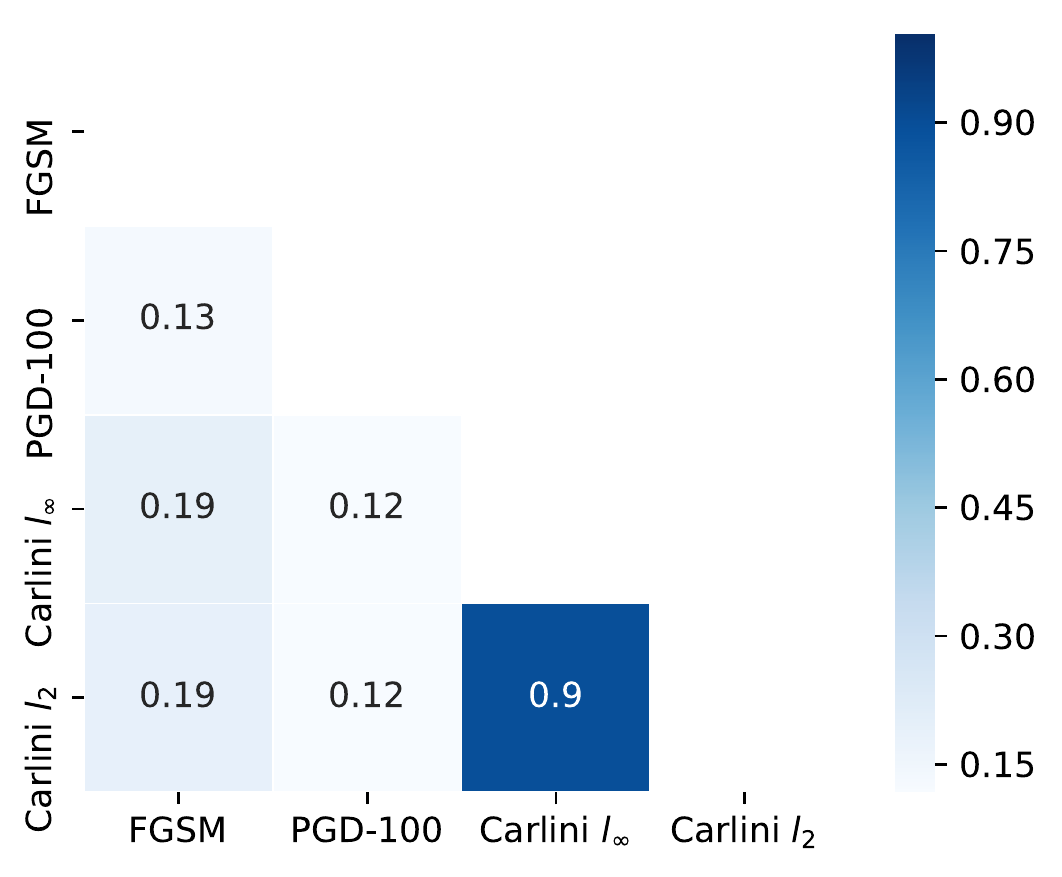}
        \caption{$\epsilon=0.005$}    
        \label{fig:cmat_eps_0.005}
    \end{subfigure}
    \caption{Similarity (on a scale of [0,1]) between wrong predictions made by the model for different attacks.} 
    \label{fig:app similarity mis-classification}
\end{figure*}

\section{Visualizing spectrograms}\label{app:Visualizing spectrograms}
Figure~\ref{fig:app spectrograms} shows the mel-spectrograms of a randomly chosen utterance for different attacks at varying $\epsilon$ values. 
Here, for exploratory analysis, we increase $\epsilon$ beyond the range specified in our experiments described in the main text.
We can see that for both FGSM and PGD, the noise is visible in the mel-spectrogram for $\epsilon=0.002$. 
The signal becomes extremely noisy for $\epsilon=0.2$ (SNR drops below $-10$ dB, and PESQ score $< 1.5$).
On the other hand, for Carlini $l_\infty$ attack, the noise is almost invisible at $\epsilon=0.002$ and $\epsilon=0.02$, as also evident from the high SNR values and PESQ scores.
The noise becomes somewhat visible at $\epsilon=0.2$ where the SNR drops to $10$ dB, and the PESQ score becomes $\sim 3.4$.
%%\textcolor{red}{Please visit the project webpage (PROJECT WEBPAGE) to listen to the adversarial speech samples.}

\section{Similarity in misclassification for different attacks}
\label{app:Similarity in misclassification for different attacks}
We investigate whether different attack algorithms force the model to misclassify a particular input utterance as the same (wrong) speaker. This could possibly reveal similarity between different attack algorithms.
Figure~\ref{fig:app similarity mis-classification} shows the fraction of similarity (\ie average number of matches) between the wrong predictions made by the model for different attacks.
As evident, wrong predictions for Carlini $l_\infty$ and Carlini $l_2$ attacks are very similar ($>90\%$ similarity for all the $\epsilon$ values), possibly because the inherent strategy of the Carlini attack remains the same in the two variants.
The similarity between FGSM and the two Carlini attacks is also noticeable.
More interestingly, the similarity scores tend to decrease when $\epsilon$ increases.
We hypothesize that a low $\epsilon$ constrains the attack algorithm with a smaller space for perturbation, and hence, the model generally tends to wrongly predict the closest class (one that causes the most confusion).
On the other hand, a high $\epsilon$ opens up a lot more allowed space for the perturbation, and hence, the similarity between the wrong predictions tends to decrease.

\end{appendices}

\end{document}